\documentclass[showpacs,prb,floatfix,twocolumn,amsmath,a4]{revtex4}
\usepackage{graphicx}
\usepackage{color}
\usepackage[textsize=tiny]{todonotes}

\usepackage[latin1]{inputenc}

\usepackage{amsmath}
\usepackage{amsfonts}
\usepackage{amssymb}

\renewcommand{\b }{\mathbf }
\newcommand{\m}[1]{\underline{\underline{#1}}}

\begin{document}

\title{Critical slowing down near the multiferroic phase transition in MnWO$_4$}

\author{D. Niermann$^1$}
\author{C.P. Grams$^1$}
\author{P.~Becker$^{2}$}
\author{L.~Bohat\'y$^{2}$}
\author{H.~Schenck$^{3}$}
\author{J.~Hemberger$^{1}$}
\email[Corresponding author:~]{hemberger@ph2.uni-koeln.de}

\affiliation{$^1$II.\ Physikalisches Institut, Universit\"at zu K\"oln, Z\"ulpicher Str.\ 77, D-50937 K\"oln}
\affiliation{$^2$Institut f\"ur Kristallographie, Universit\"at zu K\"oln, Greinstr.\ 6, D-50939 K\"oln, Germany}
\affiliation{$^3$Institut f\"ur Theoretische Physik, Universit\"at zu K\"oln, Z\"ulpicher Str.\ 77, D-50937 K\"oln}

\begin{abstract}
By using broadband dielectric spectroscopy in the radiofrequency and microwave range we studied the magnetoelectric dynamics in the
multiferroic chiral antiferromagnet MnWO$_4$.
Above the multiferroic phase transition at $T_{N2} \approx 12.6$\,K we observe a critical slowing down of the 
corresponding magnetoelectric fluctuations resembling the soft-mode behavior in canonical ferroelectrics. This electric field driven excitation carries much less spectral weight than ordinary phonon modes. Also the critical slowing down of this mode scales with an exponent larger than one which is expected for magnetic second order phase transition scenarios. Therefore the investigated dynamics have to be interpreted as the softening of an electrically active magnetic excitation, an electromagnon. 
\end{abstract}

\date{\today}


\maketitle

The field of magnetoelectric multiferroics has continuously grown and meanwhile constitutes an established research area in condensed matter physics \cite{Khomskii09,Spaldin05}. In single phase multiferroics magnetic and electric order coexist and often, in the case of magnetoelectric coupling, form a common order parameter. One meanwhile well studied mechanism for intrinsic magnetoelectric coupling is based on the inverse Dzyaloshinskii-Moriya (DM) interaction in partially frustrated spiral magnets \cite{Cheong07,Tokura09}. This scenario of ferroelectricity, which is driven directly by complex magnetism, provides the underlying mechanism e.g.\ for the family of orthorhombic rare earth manganites $R$MnO$_3$, from which the renewed interest in magnetoelectric multiferroicity started about 10 years ago \cite{Kimura03, Hemberger07}. In these compounds a non-collinear cycloidal spin-structure is formed, which leads to a coherent distortion of the Mn-O-Mn bonds resulting in a ferroelectric polarization. 
However, the formation of a multiferroic phase with a complex, hybrid order parameter raises questions concerning the dynamics of corresponding fluctuations.

The elementary excitations within an ordered multiferroic phase are called electromagnons and were detected, e.g., in orthorhombic manganites \cite{Pimenov06,Katsura07,Senff07}. Electromagnons combine the  excitation of the magnetic structure with a polar lattice distortion and can be regarded as magnons excitable by an electric field. These excitations are typically located at tera\-hertz frequencies below the ordinary phonon region. 
Moreover, this type of excitation may be associated with the symmetry breaking Goldstone-mode of the  multiferroic order \cite{Shuvaev10,Pailhes09,Katsura07}. This would imply that similar to generic ferroelectrics the softening of this mode has to be expected near the second order phase transition. Thus, such a soft-mode scenario also should influence the dynamic response of fluctuations above the onset of static multiferroic order and implies electromagnon excitations even above the multiferroic phase \cite{Pimenov08,Sushkov08}.

\begin{figure}
\centerline{\includegraphics[width=0.6\columnwidth,angle=0]{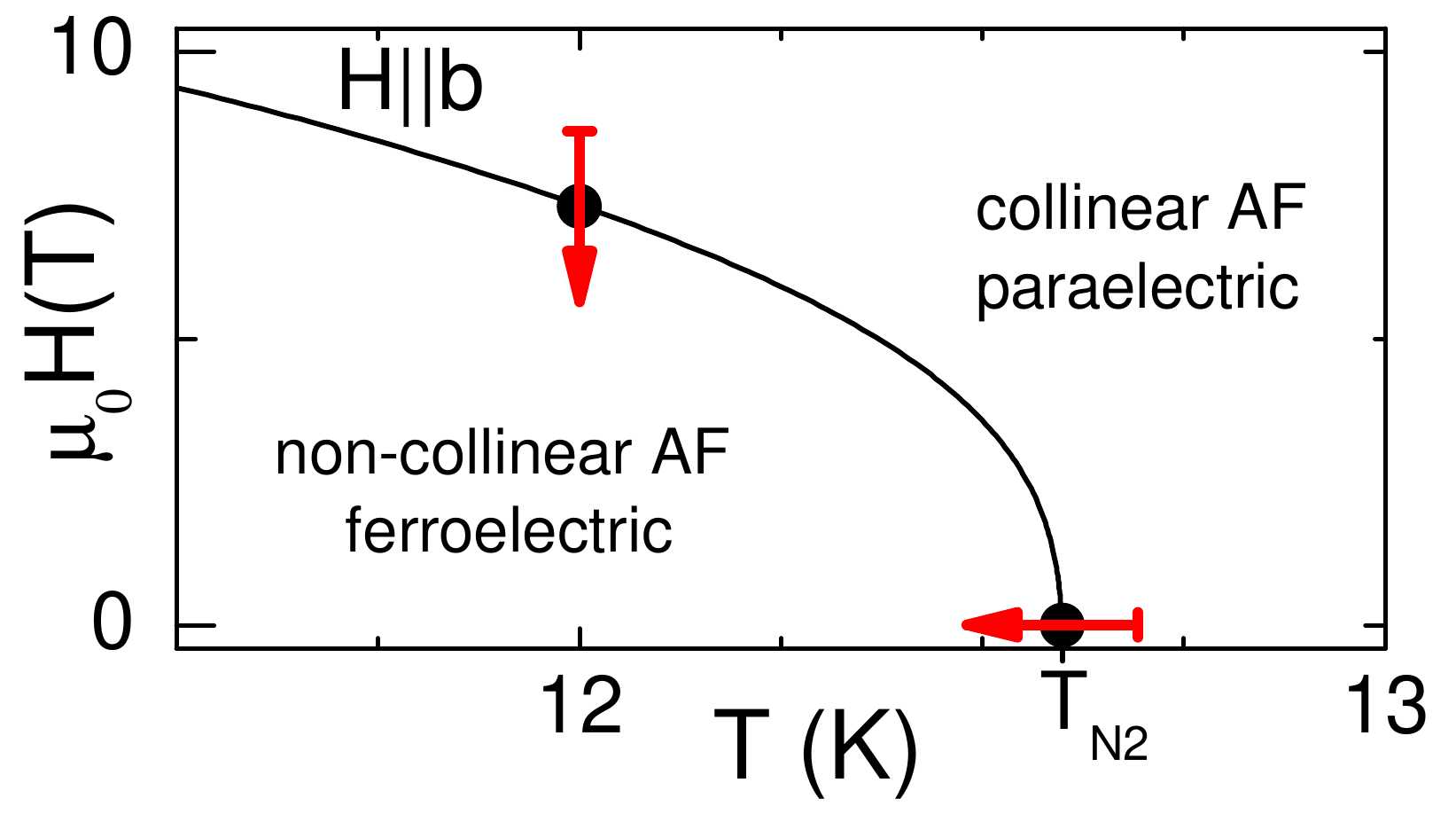}}
\caption{
(Color online) Schematic $(H,T)$-diagram for the multiferroic phase in MnWO$_4$ (see also Refs.~\onlinecite{Taniguchi06,Arkenbout06}). The red arrows denote the two ways used in this study to approach the phase transition, namely temperature, driven in zero-field and magnetic-field driven at $T=12$~K. 
}
\label{figHTdiagram}
\end{figure}

MnWO$_4$ belongs to the above described type of compounds, in which the ferroelectricity is driven by the onset of chiral spin order via the inverse DM interaction \cite{Heyer06,Taniguchi06,Arkenbout06}.
The spin-system undergoes a sequence of phase transitions: Cooling down from the paramagnetic high temperature phase at $T_{N3}=13.5$\,K an incommensurate, collinear antiferromagnetic, sinusoidal spin-density wave with an easy spin-axis within the $ac$-plane emerges. At $T_{N2}\approx 12.6$~K a second order transition into a non-collinear phase with cycloidal spin-structure.
This is the actual multiferroic phase in which the spatial inversion symmetry is broken and a finite electric polarization along the $b$-direction is established  \cite{Heyer06,Taniguchi06,Arkenbout06}. 
The ferroelectricity is lost again below a first-order phase transition at $T_{N1}\approx 7.6$~K where the system realizes a commensurate and collinear spin-order. 
Thus the multiferroic phase 
lies in the temperature range $T_{N1}=7.6$\,K\,$<T<T_{N2}=12.6$\,K, embedded between a paraelectric phase with collinear sinusoidal spin order at higher $T$ and a paraelectric phase with collinear antiferromagnetic spin order at lower $T$. It is also possible to destabilize the cycloidal spin-structure with an external magnetic field along the $b$-axis\cite{Taniguchi06,Arkenbout06}. The field continuously bends the spins out of the ac-plane, transforming the planar cycloidal spin-structure into a conical one  until the sinusoidal phase is restored. Thus not only temperature but also magnetic field can act as  control parameter for the multiferroic order. The corresponding $(H,T)$-diagram is illustrated in Fig.~\ref{figHTdiagram}.   
In this article we report on spectroscopic investigations of the complex permittivity in high quality MnWO$_4$ single crystals for frequencies from 30~MHz to 1.8~GHz in order to shed light on the dynamical dielectric response of the magnetoelectric fluctuations on approaching the multiferroic phase transition.

The single crystals of MnWO$_4$ used for sample preparation were grown from the melt using a top-seeded growth technique\cite{Becker07}.
The crystals were structurally and magnetically characterized revealing the expected monoclinic space group P2/c \cite{Macavei93} and the above described sequence of phase transitions at $T_{N1}\approx 7.6$\,K, $T_{N2}\approx12.6$\,K, and $T_{N3}\approx13.5$\,K \cite{Heyer06,Taniguchi06,Arkenbout06}.
The complex, frequency dependent dielectric response $\varepsilon^*(f)$ was measured on plate-shaped samples using a home-made coaxial-line inset in a commercial $^4$He-flow magneto-cryostat ({\sc Quantum-Design PPMS}) and evaluated from the complex transmission coefficient ($S_{12}^*$) as determined via a vector network analyzer ({\sc Rohde \& Schwarz}). All  measurements were performed with the electric field along the crystallographic $b$-axis, the direction in which the spontaneous ferroelectric moment points in zero external magnetic field \cite{Taniguchi06,Arkenbout06}. The measurements of the complex permittivity were carried out in a frequency range from 10~MHz to 2~GHz with driving-fields of the order $E_{ac} \leq 1$\,V/mm. 
In all cases, the contacts were applied to the plate-shaped single-crystal samples using silver paint in sandwich geometry with a typical electrode area of $A\approx 2$\,mm$^2$ and a sample thickness of $d\approx0.3$\,mm.
The uncertainty in the determination of the exact sample geometry 
together with additional (but constant) contributions of stray capacitances results in an uncertainty in the absolute values for the permittivity of up to 30\,\%.

Usually, in multiferroic compounds measurements of the dielectric permittivity are performed at one single frequency, e.g.\ 1\,kHz or 10\,kHz, in order to determine the transition temperatures. \cite{Kimura03,Taniguchi06,Arkenbout06,Heyer06} In proper ferroelectrics one expects divergent behavior for the quasi-static permittivity on approaching the phase transition, even though $\varepsilon'(T)$ always will stay finite due to damping \cite{Blinc74}. In improper ferroelectrics, such as the regarded multiferroics, where the ferroelectric component acts only as a secondary order parameter, the "divergence" of $\varepsilon'(T)$ is only mimicked by a small anomaly, which usually is even smaller than dielectric background $\varepsilon_\infty$, which results from the phononic polarizability of the lattice and is non-dispersive in the sub-phononic frequency range. In MnWO$_4$ this dielectric background is $\varepsilon_\infty\approx 15$, while the additional contribution due to the onset of ferroelectricity at $T_{N2}$ stays below $\Delta\varepsilon \leq 1$ \cite{Heyer06,Taniguchi06,Arkenbout06}.  
The small weight of the anomaly at the transition compared to the dielectric background reflects, that the ferroelectric moment of the multiferroic phase is nearly four orders of magnitude smaller than in ordinary ferroelectrics \cite{Blinc74}. 

\begin{figure}
\centerline{\includegraphics[width=0.7\columnwidth,angle=0]{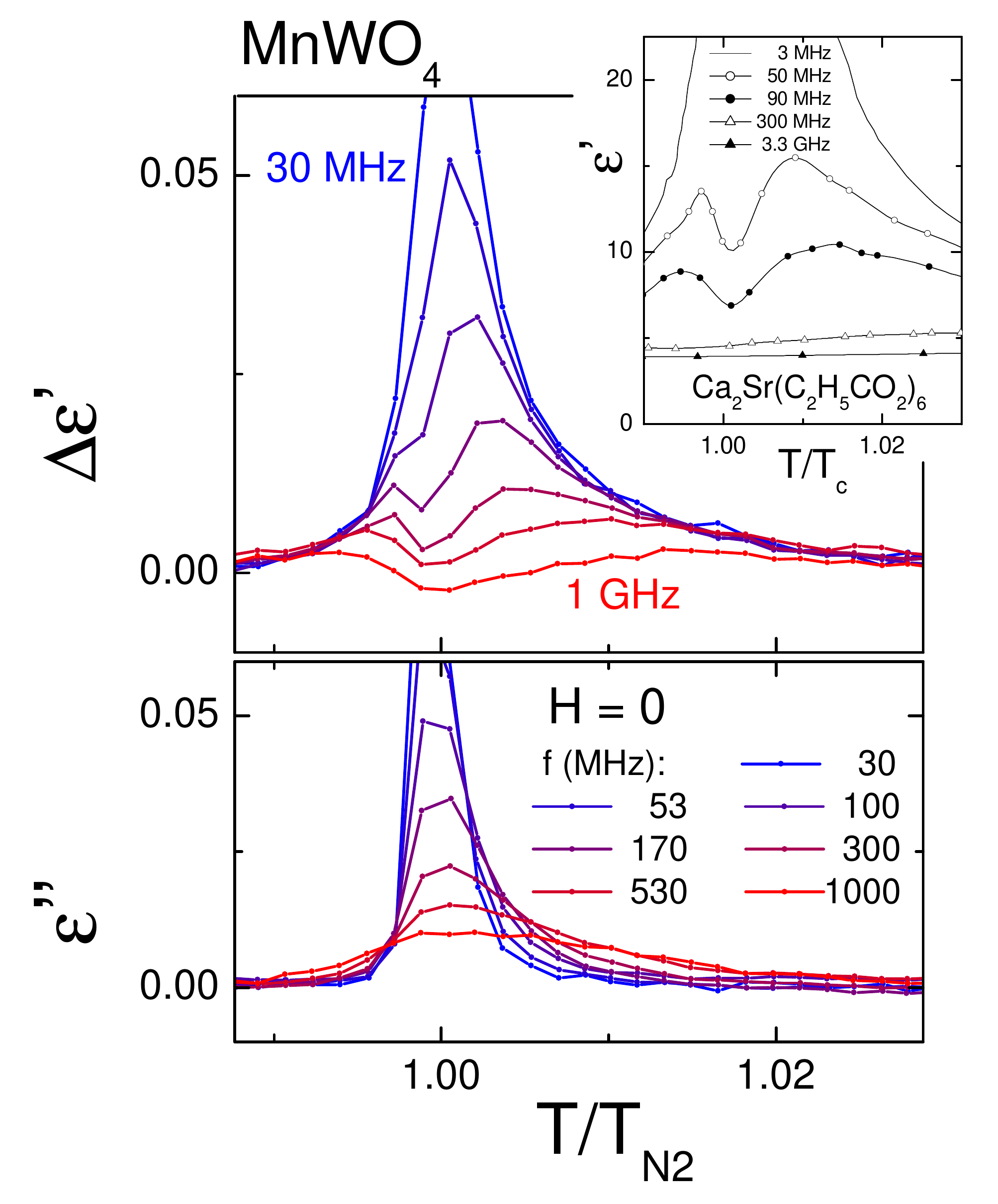}}
\caption{
(Color online) Real part $\Delta \varepsilon'(T)$ (only the ferroelectric contribution, upper panel) and imaginary part $\varepsilon''(T)$ (lower panel) of the complex dielectric permittivity for frequencies
between 30~MHz and 1~GHz.
around the multiferroic phase transition. For comparison the inset shows similar measurements in a canonical proper ferroelectric (Ca-Sr-propionate) taken from \cite{Blinc74}.
Note that the scales differ by four orders of magnitude due to the small ferroelectric moment in MnWO$_4$.
}
\label{figHF}
\end{figure}    

\begin{figure}
\centerline{\includegraphics[width=0.6\columnwidth,angle=0]{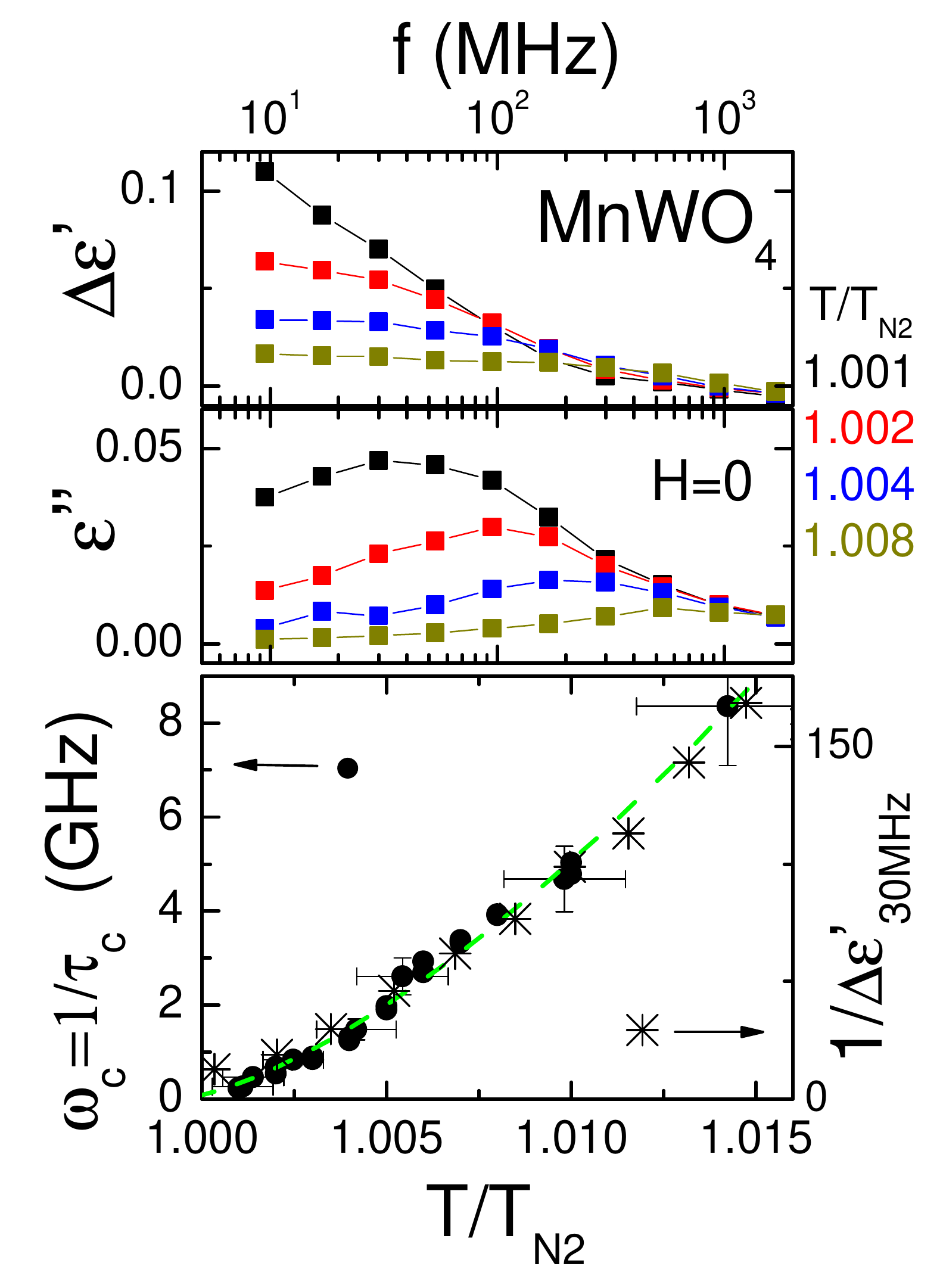}}
\caption{
(Color online) Upper frames: Spectra of the real and imaginary part of the complex permittivity $\varepsilon(f)^*$ for exemplary temperatures slightly above $T_{N2}$.
Solid lines are guide to eyes. 
Lower frame: Critical temperature dependence of the relaxation rate ($\bullet$, left scale) $\omega_c(T)=2\pi f_p$ as extracted from the position $f_p$ of the peaks in the 
loss spectra (upper frame). The scattered line (green) is a fit using the 
expression 
$f_p\propto \left[{(T-T_{N2)}}/{T_{N2}}\right]^{\nu z}$ 
with $\nu z\approx 1.3$. 
Right scale ($\ast$): The inverse of the real part of the permittivity measured at $f=30$~MHz.
}
\label{figTauSpectrum}
\end{figure}

Fig.~\ref{figHF} shows  $\varepsilon*(T)$ around $T_{N2}$ for frequencies $f$
\footnote{In this publication the frequency is denoted by $f$ to avoid confusion with the critical exponent $\nu$ as used by several authors.} 
between 30\,MHz and 1\,GHz. 
The real part is shown as $\Delta \varepsilon'(T)= \varepsilon'(T) - \varepsilon_\infty$ after subtracting the dielectric background, as gained from the dispersionless region well above the transition for each frequency. 
This subtraction allows to make use of the high relative resolution of the measurement system.
For lower frequencies $\Delta\varepsilon'(T)$ shows the critical, i.e. divergent, behavior expected for a second order phase transition. A divergent static susceptibility results from a diverging spatial correlation length. 
However, to capture the behavior of the dynamical susceptibility the increasing correlation time of the fluctuations has to be taken into account as well.
These time scales should show up in the experiment as dispersive dielectric response. 
Indeed we find dispersion at temperatures above the transition in the higher frequency measurements of $\varepsilon'(T)$ (upper panel of Fig.~\ref{figHF}) which break away from the quasi-static divergent behavior when the experimental frequency of the stimulus meets the fluctuation rate. Correspondingly, 
contributions to the dielectric loss $\varepsilon''(T,f)$ appear, as monitored in the lower panel of Fig.~\ref{figHF}.  
These dispersive features can be parametrized using an effective relaxation time $\tau_c$, which represents the fluctuation lifetime and which has to increase on approaching the continuous phase transition. The corresponding Debye-type relaxation spectra, i.e.\ a step in $\varepsilon'(f)$ accompanied by a peak in $\varepsilon''(f)$, are displayed for different temperatures in the upper panels of Fig.~\ref{figTauSpectrum}. When the corresponding relaxation rate is met by the angular excitation frequency ($1/\tau_c=\omega=2\pi f$), the regarded loss features occur and the permittivity deviates from its static limit. 
At $T_{N2}$, where $\tau_c(T)$ is largest, the higher frequency response forms a characteristic minimun in $\varepsilon'(T)$, which can be taken as fingerprint for such a {\em critical slowing down} scenario \cite{Blinc74,Hemberger96}.
The inset of Fig.~\ref{figHF} displays textbook-like data for a proper ferroelectric material \cite{Blinc74} and hereby demonstrates a very good qualitative agreement for the case of multiferroic MnWO$_4$. 

It should be noted 
that these findings 
can clearly be distinguished from any relaxation phenomena driven by defects or quenched disorder. 
Electrical random fields 
are known to lead to the formation of relaxor ferroelectric phases \cite{Kleeman07, Bednorz84} and localized carriers may lead to trapped polarons, coupled to the magnetic ordering\cite{Goto05,Schrettle09}.
The experimental signature of such effects may indeed may be broad relaxation features  but none of these pinning induced freezing scenarios is able to explain the observed minimum $\varepsilon'(T)$, which results from the critical ``softening'' and ``re-hardening'' of the fluctuation rate going trough a continuous phase transition. 
However, below the transition in the ordered multiferroic phase these critical dynamics can be masked by contributions of domains and domain-walls, respectively  \cite{Kagawa09,Niermann14}. Therefore, we will restrict the evaluation of the dynamics to the region $T>T_{N2}$ in the following.

Experimentally the critical time scale is defined by the step in $\Delta\varepsilon'(f)$ and the peak 
in the corresponding loss spectra $\varepsilon''(f)$ (see upper panels of Fig.~\ref{figTauSpectrum}).
This position of the loss maximum at different temperatures determines the relaxation rate $\omega_c=1/\tau_c = 2\pi f_{p}$.
Even though the data are limited by experimental resolution due to the relatively small dielectric contribution, it is possible to evaluate them with respect to an effective peak frequency $f_p(T)$.
The result is shown in the lower panel of Fig.~\ref{figTauSpectrum}:
The relaxation rate (denoting the inverse correlation time) continuously diminishes approaching the multiferroic transition.

The relationship between the critical slowing down of the fluctuation dynamics and the critical increase of the quai-static permittivity towards the transition can be understood in analogy to the softening of a polar lattice mode in proper ferroelectrics, as described by the Lyddane-Sachs-Teller (LST) relation $\varepsilon_s \propto \omega_0^{-2}$, where $\omega_0$ is the angular eigenfrequency of the undamped mode and $\varepsilon_s$ denotes the static permittivity \cite{Blinc74}. Considering a constant damping $\Gamma$ such a softening mode becomes overdamped in the vicinity of a phase transition and can be described via a critical timescale $\tau_c\approx \Gamma/\omega_0^2$. So, the LST-relation for the overdamped case transforms to $\varepsilon_s \propto \tau_c$, which corresponds to the mean field result for dynamic critical scaling \cite{Hohenberg77}. The validity of this scaling is demonstrated by the temperature dependence of the inverse permittivity contribution at low frequency $1/\Delta{\varepsilon'_s}\propto 1/\tau_c$ as displayed in the lower panel of Fig.~\ref{figTauSpectrum} (right scale, $\ast$).

The relaxation rate $\omega_c = 1/\tau_c$ decreases continuously on approaching the multiferroic transition, resembling the behavior of an overdamped soft-mode, like in proper ferroelectrics \cite{Blinc74}. However, in the present case the driving order parameter is of magnetic origin and the corresponding excitations are of electromagnonic nature \cite{Katsura07}. Therefore, no distinct softening of a phonon mode may be observed but only the transfer of much less optical weight from a phonon to a softening electromagnon may occur \cite{Moeller14}. 
Even so it is difficult to evaluate the optical weight $\sigma'(\omega)d\omega=\varepsilon_0\int \omega \varepsilon''(\omega)d\omega$ from our dielectric data (middle frame of Fig.\,\ref{figTauSpectrum}) due to the limited 
frequency span, one can estimate a lower boundary which would be compatible with an optical weight six orders of magnitude smaller than that for the lowest optical phonon in MnWO$_4$ \cite{Choi10,Moeller14} 
or for the dominant electromagnon excitation in the cycloidal phase of TbMnO$_3$\cite{Pimenov06}, which however, is mediated via a different coupling mechanism \cite{Aguilar09}. 
This meets the findings of an only very small spectral weight transfer from phonons to the electromagnon investigated by IR-spectroscopy \cite{Moeller14}.
Thus we claim that MnWO$_4$, compared to e.g.\ TbMnO$_3$, possesses a significantly weaker magnetoelectric coupling and therefore the electro-magnonic excitation in electric field possesses only small spectral weight.

We tried to parameterize the critical slowing down scenario using the expression  $f_p\propto1/\tau_c\propto[(T-T_{N2})/T_{N2}]^{\gamma}$. Fitting the data yields a critical exponent of $\gamma\approx 1.3$. While the exact value should not be overrated due to the measurement uncertainty, it is remarkable that it differs from the canonical expectation for proper ferroelectrics where $\gamma = 1$. This again points to the predominantly magnetic nature of the multiferroic transition.  
Using the Ginzburg-Landau expansion for MnWO$_4$ done by Tol\'{e}dano et.~al.~\cite{Toledano10}, we can identify the transition at $T_{N2}$ as being in the 3D-Ising universality class, in agreement with the literature \cite{Matityahu12,Harris08}. The critical slowing down is therefore described by an overdamped Ising-order parameter, known as kinetic Ising model or Model A \cite{Hohenberg77}. The dynamical exponent $z$ characteristic for these models describes the critical behavior of the relaxation time $\tau_c$ and is related to the measured exponent via $\gamma = \nu z$, where $\nu$ is the critical exponent of the correlation length.\cite{Hohenberg77}
The exponent $z$ has been calculated to 4-loop order by Prudnikov et.~al.~\cite{Prudnikov97} as $z = 2.017$. An overview over different techniques and comparison to numerical values can be found in Folk et.~al.\cite{Folk06}.
With the exponent $\nu \approx 0.6304$ calculated by Guida and Zinn-Justin \cite{Guida98} we get $\nu z \approx 1.272$, which is in very good agreement with our experimental findings. A detailed derivation of the exponent $\nu z$ can be found in the supplement material \cite{supplement}.
It is in accordance to canonical ferroelectrics that for the regarded frequency range up to roughly 1~GHz the measured loss features appear in a temperature regime of only 2\,\% of $T_{N2}$ above the transition. 
This agrees with the picture that the fluctuations of the magnetoelectric order parameter in MnWO$_4$ are of three-dimensional nature like in ferroelectrics and thus do not extend to temperatures high above the ordering transition as it can be the case, e.g., for low-dimensional magnets.

\begin{figure}
 \centerline{\includegraphics[width=0.7\columnwidth,angle=0]{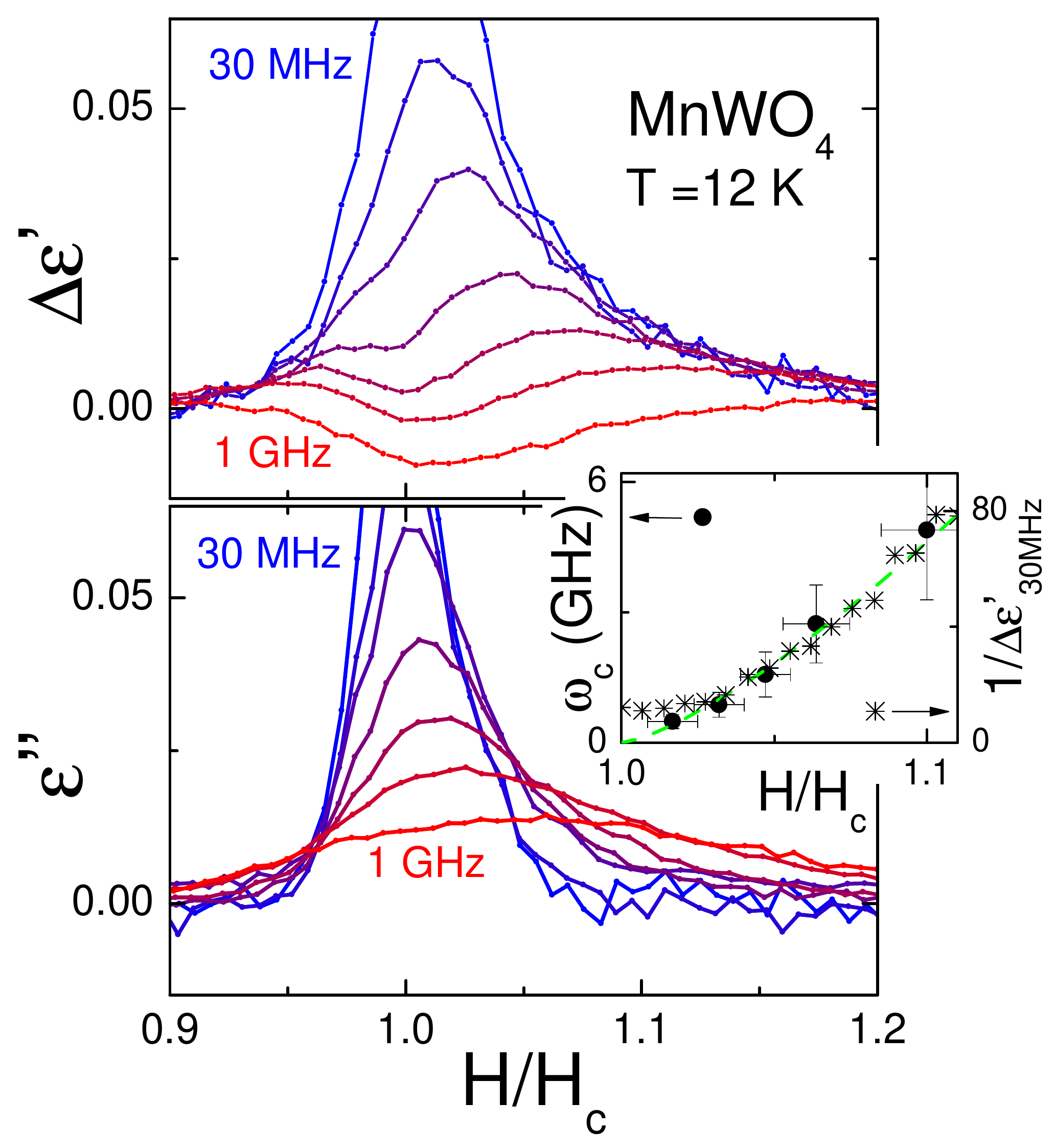}}
\caption{
(Color online) Real part $\Delta \varepsilon'(T)$ (upper frame) and imaginary part $\varepsilon''(T)$ (lower frame) of the complex 
permittivity for frequencies between 30\,MHz and 1\,GHz (see Fig.~\ref{figHF}) as function of the magnetic field at $T=12$\,K. The inset shows the critical magnetic field dependence of the relaxation rate $\omega_c=1/\tau_c$ (left scale, $\bullet$). The solid line is a fit due to $f_p \propto  [({H-H_{c}})/{H_{c}}]^{\nu z_\text{\tiny{H}}}$ at $H_c\approx 7.3$~T with 
$\nu z_\text{\tiny{H}}\approx 1.3$. The right scale of the inset ($\ast$) refers to the inverse of $\Delta \varepsilon'(T)$
measured at $f=30$~MHz.
}
\label{figeps(H)}
\end{figure}

The critical slowing down scenario of the fluctuation dynamics can also be demonstrated for the magnetic-field-driven transition. Applying a magnetic field along the crystallographic $b$-axis weakens the chiral magnetic structure and shifts the phase boundary towards lower temperatures \cite{Arkenbout06,Taniguchi06}. 
At $T=12$\,K a small variation of the external magnetic field around the critical value $\mu_0 H_{c}= 7.3$\,T allows to change continuously between the chiral multiferroic and the sinusoidal collinear magnetic phase. 
Fig.~\ref{figeps(H)} shows data of the complex permittivity for similar frequencies, as shown in Fig.~\ref{figHF}, but this time as function of the normalized magnetic field. The results denote a qualitatively similar behavior: A characteristic minimum for the higher frequencies in $\varepsilon'$ and the occurrence of dielectric loss already at fields above $H_c$ probing the dissipative stimulation of fluctuations. 
Analogous to the case of temperature as control parameter, the critical fluctuation rates, as extracted from the loss maxima (inset of Fig.~\ref{figeps(H)}) were modeled with a function $f_p \propto 1/\tau_c \propto [({H-H_{c}})/{H_{c}}]^{\nu z_\text{\tiny{H}}}$. Again we find a critical exponent larger than unity ($\nu z_\text{\tiny{H}}\approx1.3$).

Summarizing, we demonstrated for the first time the softening of a magnetic mode at a continuous multiferroic phase transition via the dynamical magnetoelectric response at GHz frequencies.
Above the continuous multiferroic phase transition in MnWO$_4$ a critical slowing down of the ferroelectric fluctuations can be monitored via the evaluation of relaxational permittivity contributions in the GHz range for $E||b$, the crystallographic axis along which spontaneous polarization forms below $T_{N2}$. The real part of the permittivity $\varepsilon'(T)$ shows a characteristic minimum due to critical damping at the transition temperature, which is known from proper ferroelectrics exhibiting much higher values of permittivity or polarization. The corresponding loss spectra $\varepsilon''(\nu)$  above $T_{N2}$ show characteristic maxima from which fluctuation rates can be determined. This scenario can be understood as the softening of an overdamped polar mode on approaching the multiferroic phase transition. However, the small spectral weight of this electric-field driven excitation suggests an (electro-)magnonic nature resulting from the only weak magnetoelectric coupling. The mode softening not only can be driven via temperature but also via an external magnetic field, in both cases yielding a critical exponent $\nu z$ larger than unity. This points towards the predominantly magnetic origin of the underlying fluctuations, which are well described theoretically by a single overdamped magnetic 3D-Ising order-parameter.
These results suggest to use other dynamical probes coupling directly to the magnetic fluctuations like spin-echo or inelastic neutron investigations. 
It will be interesting to compare these dynamical characteristics to multiferroics possessing different magnetoelectric coupling mechanisms.

This work has been funded by the DFG through SFB608 (Cologne) and research grant HE-3219/2-1 as well as the Institutional Strategy of the University of Cologne within the German Excellence Initiative.


\newpage
\subsection{Appendix: 
Supplementary material concerning the dynamical exponent $\nu z$ of the temperature dependent relaxation time $\tau_c$ of the polarization $P_y$ }

The magnetic structures of MnWO$_4$ compatible with the P2/c space group can be understood using Bertaut's group theoretical considerations, where the irreducible representations of the space group determine the 
coupling of the Fourier components of the magnetic moments \cite{Lautenschlaeger93, Bertaut68}. 
In the case of incommensurable Phases AF3 and AF2 with $\mathbf{k}_{\text{inc}}=(-0.214,1/2,0.457)$ \cite{Lautenschlaeger93} there are two associated bidimensional irreducible corepresentations\cite{Toledano10} corresponding to two complex order-parameters
$\tilde{S}_1 = S_1 e^{\imath \theta_1} = s_1 + \imath \bar{s}_1$ and $\tilde{S}_2 = S_2 e^{\imath \theta_2} = s_2 + \imath \bar{s}_2$.
Each complex order parameter is related to the Fourier coefficients of a sinusoidal wave of the magnetic moments of the Mn$^{2+}$, encoding amplitude and phase, where both waves are perpendicular to each other \cite{Lautenschlaeger93,Harris07,Mostovoy06}.

Using the space group symmetry, Tol\'{e}dano et.~al.~\cite{Toledano10} calculated the Landau expansion of the free energy $\Phi$ of MnWO$_4$ in $\tilde{S}_1$ and $\tilde{S}_2$ as
\begin{align}
 \Phi &= \Phi_0(T) + \Phi^D + \frac{\alpha_1}{2} S_1^2 + \frac{\beta_1}{4} S_1^4 + \frac{\alpha_2}{2} S_2^2 + \frac{\beta_2}{4} S_2^4 \nonumber \\
  &\quad + \frac{\gamma_1}{2} S_1^2 S_2^2 \cos(2 \varphi) + \frac{\gamma_2}{4} S_1^4 S_2^4 \cos^2(2\varphi) + \dots 
\end{align}
with a coupling to the polarization $P_y$ via
\begin{equation}
 \Phi^D = \delta P_y S_1 S_2 \sin(\varphi) + \frac{1}{2 \varepsilon_{yy}^0} P_y^2
\end{equation}
where $\varphi= \theta_1 - \theta_2$ \cite{Toledano10}, which they successfully employed in mean-field theory to describe the square-root temperature dependence of the polarization and the Cure-Weiss-type behavior 
of the dielectric permittivity $\varepsilon^0_{yy}$\cite{Toledano10} as measured by Taniguchi et.~al \cite{Taniguchi06,Toledano10}. Above $T_{N3}$ in this picture both order parameters are zero on average, corresponding
to the paramagnetic phase P \cite{Toledano10}. The transition from the paramagnetic phase P to the incommensurate magnetic phase AF3 happens when $\tilde{S}_2$ becomes critical at $T_{N3}$, resulting in a non-vanishing 
equilibrium value $\tilde{S}^e_2 \neq 0$ \cite{Toledano10}, which corresponds to a sinusoidal wave structure in the magnetic moments of the Mn$^{2+}$ as described by Mostovoy \cite{Mostovoy06} and does not induce 
a Polarization. Moving to $T_{N2}$, where the transition to the incommensurable elliptical spiral phase occurs, $\tilde{S}_1$ becomes critical with $\tilde{S}_e^1 \neq 0$ and the phase difference $\varphi$ is locked as $\varphi=(2n+1)\pi/2$ \cite{Toledano10, Matityahu12,Lautenschlaeger93}.
In the ``spiral formulation'' \cite{Harris07,Mostovoy06} this means the two perpendicular sinusoidal waves corresponding to the order-parameters lock phase and are non-vanishing, forming an elliptical incommensurable 
spin structure, breaking the inversion symmetry and resulting in a non-vanishing polarization.

Now we will focus in more detail on the second transition from AF3 to AF2 where the elliptical spiral phase occurs.
When approaching $T_{N2}$ the order parameter $\tilde{S}_2$ can
be considered frozen \cite{Toledano10}. Writing the free energy near $T_{N2}$ in terms of the real and imaginary part $s_1$ and $\bar{s}_1$ of $\tilde{S}_1$ we get
\begin{align}
 \Phi &= \Phi_0'(T) +  \frac{1}{2} \b p^T \, \m M \, \b p + \frac{\beta_1}{4}(s_1^2 + \bar{s}_1^2)^2 \nonumber \\
 &\quad + \frac{\gamma_2}{4} (r_1 (s_1^2 - \bar{s}_1^2) + 4 r_2 s_1 \bar{s}_1)^2
\end{align}
where we introduced the vector $\b p = (s_1,\bar{s}_1,P_y)$ and the constants $r_1 = (s_2^2 - {\bar{s}_2}^2)$ and $r_2 = s_2 \bar{s}_2$. $\m M$ is given by
\begin{equation}
 \m M = \begin{pmatrix} \alpha_1 + \gamma_1 r_1 & 2 \gamma_1 r_2 &  - \delta \bar{s}_2 \\ 
  2 \gamma_1 r_2 & \alpha_1 - \gamma_1 r_1 &   \delta s_2  \\
  - \delta \bar{s}_2 &  \delta s_2  & \frac{1}{\epsilon_{yy}^0}
  \end{pmatrix}
\end{equation}
In order to classify the type of transition, we want to rotate into the eigensystem of $\m M$ to see how many order parameters components become critical simultaneously. Introducing the new 
components $\b u = (u_1,u_2,u_3)$ with 
\begin{equation}
 \b p^T \m M \b p = \sum_i \lambda_i u_i^2
\end{equation}
where $\{\lambda_i\}$ are the eigenvalues of $\m M$, we
find that only $u_1$ becomes critical at the AF3 $\rightarrow$ AF2 transition for the physically realized case. Since $P_y$ can be written as a linear combination of the $\{u_i\}$ and only $u_1$ is critical at $T_{N2}$
the critical behavior of $P_y$ is determined by only one magnetic order-parameter component $u_1$ and lies in the 3D-Ising universality class. This agrees with the discussion by Matityahu et.~al. \cite{Matityahu12} and reference
therein Harris et.~al. \cite{Harris08}, where the examined material is NVO (Ni$_3$V$_2$O$_8$) whose crystal structure is invariant under the same symmetry operations as MnWO$_4$. 

The Ginzburg-Landau formalism describes a static system. For the dynamics we have to make some assumptions about the type of equation of motion for the critical Ising order parameter $u_1$.
The experimental data already suggest overdamped behavior and adding a Gaussian noise term to model fluctuations we get an overdamped Langevin equation of motion for the order parameter.
This model for dynamical critical behavior is known in the literature as kinetic Ising model, or Model A \cite{Hohenberg77}. The dynamical exponent $z$ is defined as scaling exponent of the relaxation time $\tau_c$ as
\begin{align}
 \frac{1}{\tau_c} &=\nu_0 \left [ \frac{T - T_{N2}}{T_{N2}} \right]^{\nu z}
\end{align}
according to Hohenberg and Halperin \cite{Hohenberg77}. Here ${\nu}$ is the critical exponent of the correlation length $\xi$ of the 3D-Ising universality class. 
A good overview of the calculated exponents $z$ for the 3D-Ising model is contained in Folk et.~al. \cite{Folk06}, where the highest expansion is done by Prudnikov et.~al.~\cite{Prudnikov97} in terms of a 4-loop $\epsilon$ expansion with Pad\'{e}-Borel summation leading to
\begin{equation}
 z = 2.017
\end{equation}
which is very good agreement with the numerically found values around $z \approx 2.02$ found by \cite{Wansleben+91, Heuer93, Lauritsen+94}. With the value of $\nu$ found using resummation techniques (for $d=3$) 
done by Guida and Zinn-Justin \cite{Guida98}
\begin{equation}
 \nu = 0.6304 \pm 0.0013
\end{equation}
we get for the dynamical exponent of the relaxation time
\begin{equation}
 \nu z \approx 1.272
\end{equation}
which is 
in agreement with the measured value $\nu z \approx 1.3$. In accordance with the experimental observations this paints the picture that the critical dynamics of the Polarization are determined
by an overdamped magnetic 3D-Ising order-parameter.

\end{document}